# Indirect interaction in graphene nanostructures

N.N. Konobeeva[1], M.B. Belonenko[1,2]

[1] Volgograd State University, 400062, Volgograd, Russia

[2] Volgograd Institute of Business, Volgograd, Russia

E-mail: yana_nn@inbox.ru

In this paper, the form of the indirect interaction between local impurities in flakes and in graphene ring is analytically investigated. We calculate this characteristic in the framework of the tight-binding model for a finite nanocarbon system with periodic boundary conditions. A pronounced dependence of the bond value on the impurity position inside the graphene sample is found, which is due to quantum size effects. The influence of the flake size on the value of the indirect interaction constant is also studied.

**Introduction**

Since the discovery of graphene, interest in it and in the similar materials is constantly growing [1, 2]. This fact can be explained with the unique set of properties inherent in these materials. Thus, they can be used in many practical applications [3, 4]. Graphene samples used in modern experiments are obtained by micro-mechanical chipping from the surface of graphite. Such samples are called flakes. Note that such method of the production reduces the cost of the material. We also investigate graphene nanoribbons, interest in which has increased in recent years [5-7], for example, nanoribbons in the form of an "onion ring" [8]. Quantization along a circle (due to the requirement for the wave function periodicity) leads to the formation of a gap in the spectrum, which makes it possible to use these materials to create diodes, transistors, and other devices. The influence of impurities on the tunnel characteristics of the carbon structures contacts (nanoribbon, flake) is studied in [9-11]. It is shown that, within the framework of the proposed approach, the system is able to detect the presence of both impurities and defects in it.

But, one does not consider another important problem related to the use of graphene and its derivatives in spintronics, which makes it necessary to study their magnetic properties. In particular, the interactions between the magnetic moments introduced into the graphene lattice are of a great interest to researchers. One of the known mechanisms for the magnetic moments interaction is the Ruderman-Kittel-Kasuya-Yoshida (RKKI) mechanism [12-14], which is indirectly mediated by charge carriers. In [15-17] the RKKY interaction is studied within the framework of the long-wave approximation. It is shown, that the indirect interaction decreases to zero as the distance between impurity atoms increases. According these facts, the problem of investigating the exchange interaction without using the long-wave approximation in impurity flakes and rings is important and actual from a practical point of view.

**Statement of the problem**

We consider flakes of three sizes and a nanoring. The geometry is presented in Fig.1.

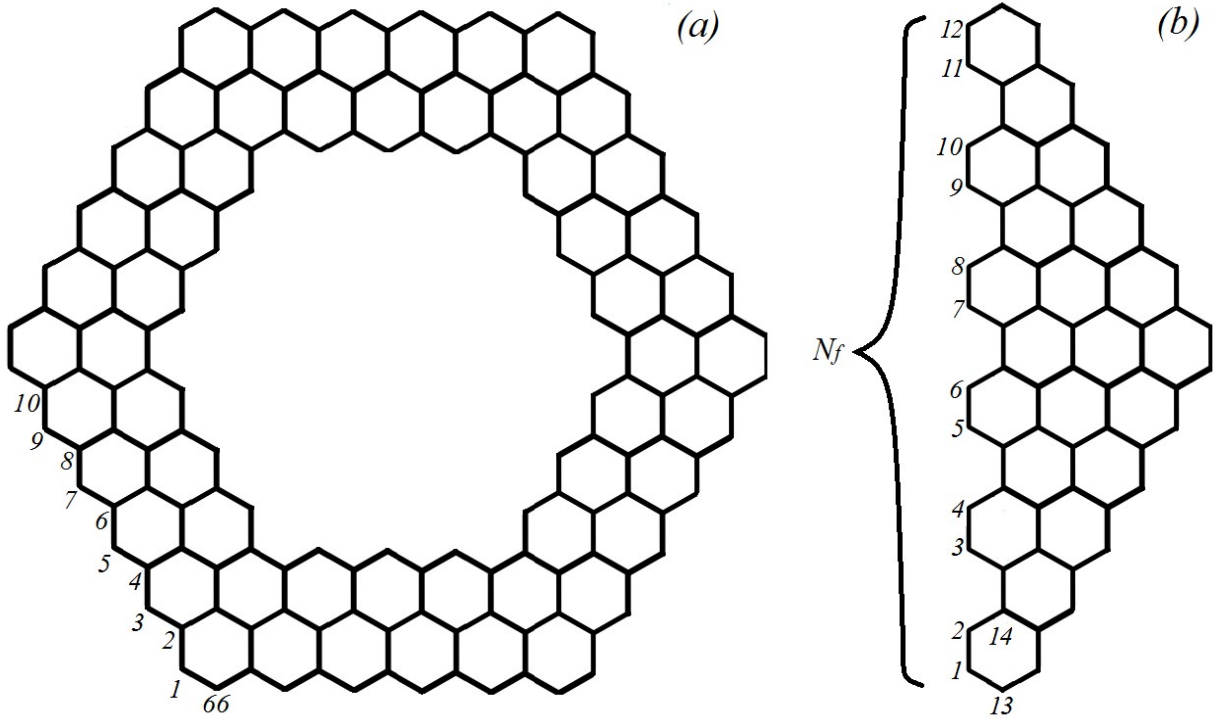

Fig.1. Geometry of the problem.

The flake electronic structure is modeled using a tight-binding Hamiltonian for $p_z$ electrons, supplemented with a Kondo impurity term [18]:

$$H = -\gamma \sum_{\langle i,j \rangle, \sigma} \left( c_{i,\sigma}^+ c_{i,\sigma} + c_{j,\sigma}^+ c_{j,\sigma} \right) + 0.5 J S_a^z \left( c_{a,\uparrow}^+ c_{a,\uparrow} - c_{a,\downarrow}^+ c_{a,\downarrow} \right) + 0.5 J S_b^z \left( c_{b,\uparrow}^+ c_{b,\uparrow} - c_{b,\downarrow}^+ c_{b,\downarrow} \right) \quad (1)$$

here $\gamma$ is the energy of the electron transfer along the interatomic bond, $\langle i,j \rangle$ notes summation over nearest neighbours in the lattice, $J$ is the contact potential, which describes the coupling between impurity spins and the electron spins at the same sites $a$ and $b$, $\sigma=\uparrow,\downarrow$ is the electron spin, $S_a^z, S_b^z = \pm 0.5$ are the impurity spins.

An example of filling the matrix form of the Hamiltonian is presented below for 4 atoms:

$$H = \begin{pmatrix} H_{11} & H_{12} \\ H_{21} & H_{22} \end{pmatrix}, H_{12} = H_{21} = 0$$

$$H_{11} = \begin{pmatrix} \frac{J}{4}+\mu & -\gamma & 0 & 0 \\ -\gamma & \mu & -\gamma & 0 \\ 0 & -\gamma & \frac{J}{4}+\mu & -\gamma \\ 0 & 0 & -\gamma & \mu \end{pmatrix}; \quad H_{22} = \begin{pmatrix} -\frac{J}{4}+\mu & -\gamma & 0 & 0 \\ -\gamma & \mu & -\gamma & 0 \\ 0 & -\gamma & -\frac{J}{4}+\mu & -\gamma \\ 0 & 0 & -\gamma & \mu \end{pmatrix} \quad (2)$$

here μ is the chemical potential, the matrix elements $H_{11}$ (when σ=↑), in which the value $\frac{J}{4}+\mu$ is found correspond to those sites with impurity. Similarly, the matrix $H_{22}$ (σ=↓) is filled.

Further, by direct diagonalization of the Hamiltonian, an effective value of the indirect interaction $J_{eff}$ is obtained.

The dependence of the effective indirect interaction on the ration $J/\gamma$ is shown in Fig. 2.

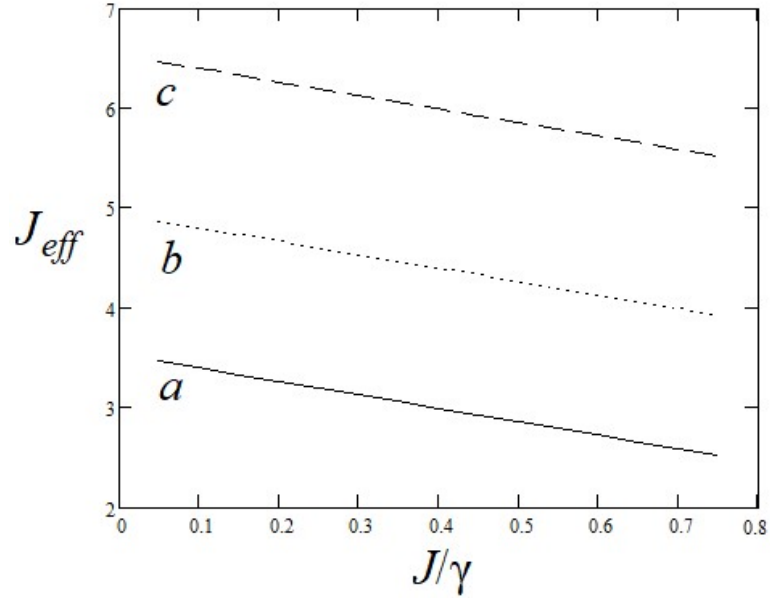

Fig.2. Dependence of the effective indirect interaction $J_{eff}$ (1 r.u.=0.1 eV) on the $J/\gamma$ for the flake with size $N_f$ (impurities are located above the lattice site, μ=0.1 eV): a) $N_f$=4; b) $N_f$=5; c) $N_f$=6.

It can be seen from Fig.2, the flake size has a great influence on the effective indirect interaction, despite of the equal distance between the impurity atoms. In this case, all three lines are parallel to each other, which indicates directly proportional relationship between the flake size $N_f$ and the magnitude of the indirect interaction.

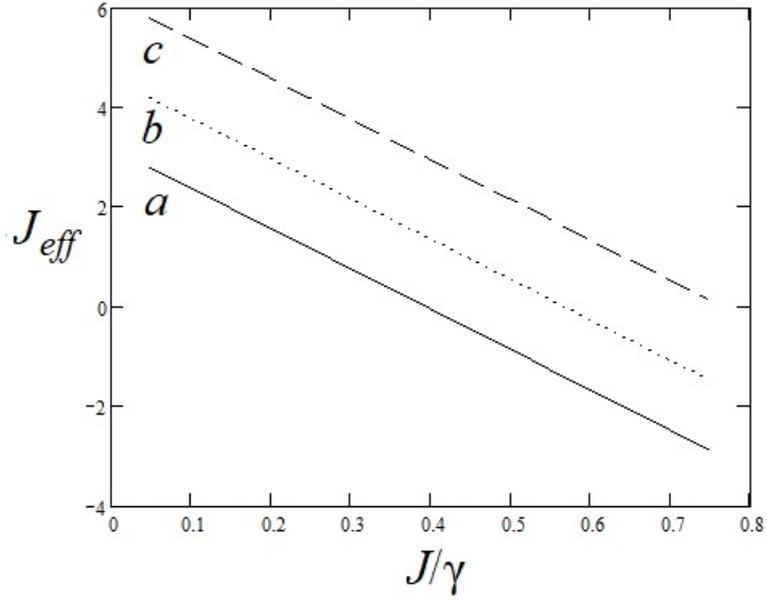

Fig.3. Dependence of the effective indirect interaction $J_{eff}$ ( 1 r.u.=0.1 eV) on the $J/\gamma$ for the flake with size $N_f$ (impurities are located above the hexagon center, $\mu$=0.1 eV): a) $N_f$=4; b) $N_f$=5; c) $N_f$=6.

We observe a similar behavior to that shown in Figure 2. Thus, the nature of the dependence of the indirect interaction on the value of $J/\gamma$ does not change, regardless of where the impurity atoms are located (above the hexagon or above a specific site).

The case corresponds to the "onion ring" is presented in Fig.4.

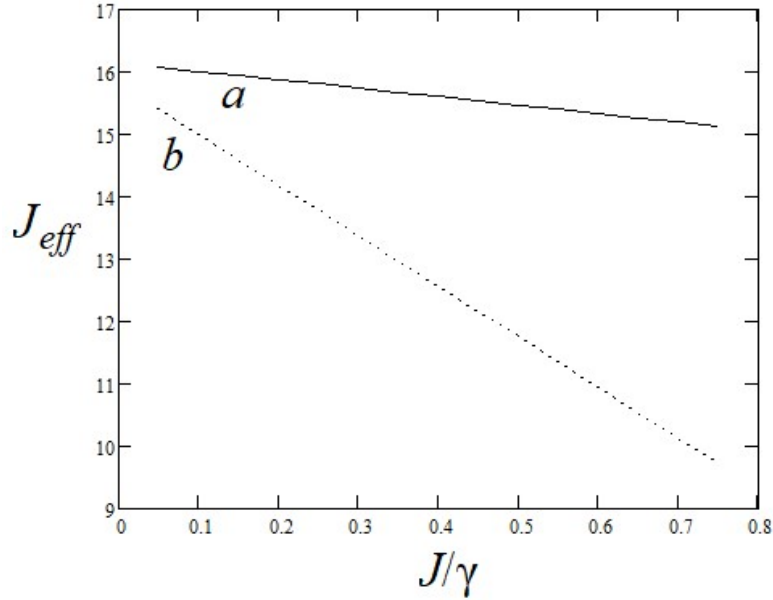

Fig.4. Dependence of the effective indirect interaction $J_{eff}$ (1 r.u.=0.1 eV) on the $J/\gamma$ for the nanoring $\mu$=0.1 eV): a) impurities are located above the lattice site; b) impurities are located above the hexagon center.

For the nanoring the constant of the indirect interaction decreases monotonically with increasing ratio $J/\gamma$. But in this case, the parallelism is not preserved for different positions of the impurity. It should be noted that here we are not talking about the distance between impurity atoms, but only about the place of adsorption of an impurity (site or hexagon).

The dependence of $J_{eff}$ on the chemical potential is shown in Fig.5.

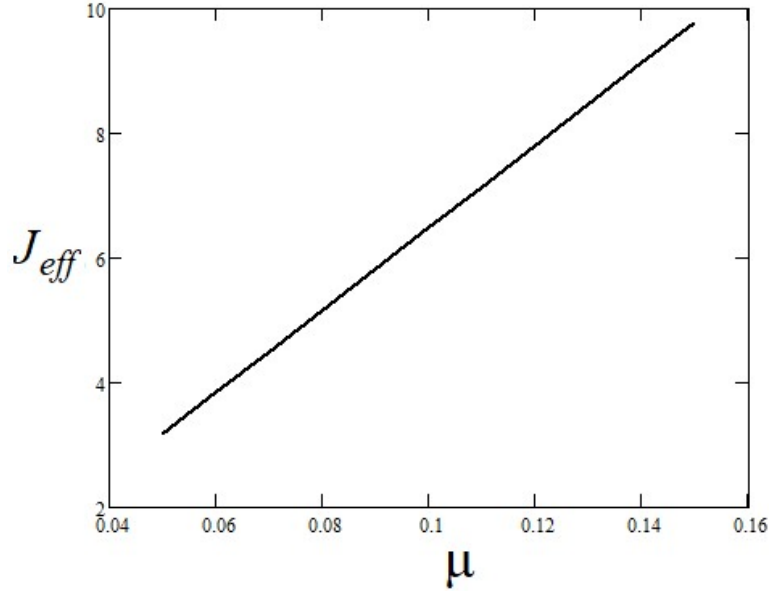

Fig.5. Dependence of the effective indirect interaction $J_{eff}$ (1 r.u.=1 eV) on the $\mu$ (1 r.u.=1 eV) for the flake with $N_f$=6 (impurities are located above the lattice site, $J/\gamma$=0.1).

When the chemical potential increases, the indirect interaction increases too. This fact associated with a change in the carrier concentration, since the magnitude of the chemical potential is precisely related to concentration.

The main results of this work may be summarized as follows:

1. We propose a method for calculating the indirect interaction constant for graphene nanoribbons by direct diagonalization of the electronic system Hamiltonian.
2. It is shown that the chemical potential of the system has the greatest influence on the magnitude of the indirect interaction.
3. We obtain that the magnitude of the indirect interaction does not depend on the distance between impurity atoms, but is determined only by the flake size.
4. In the case of nanoring, the constant $J_{eff}$ depends essentially on the adsorption place of the impurity (on the graphene ribbon atom or inside the hexagon).

**Acknowledgment**
This work was supported by the Russian Foundation for Basic Research (project no. 16-

6632-00230). Mathematical and numerical modeling is carried out within the framework of the state assignment of the Ministry of Education and Science of the Russian Federation (project no. 2.852.2017/4.6).REFERENCES

1. Novoselov K. S., Geim A.K., Morozov S.V. et al. // Science. 2004. V. 306. P. 666–669.

2. Novoselov K. S., Geim A.K., Morozov S.V. et al. // Nature. 2005. V. 438, P. 197–200.

3. G. Alymov, V. Vyurkov, V. Ryzhii, D. Svintsov // Scientific Reports. 2016. V. 6, Article number: 24654.

4. I. Shtepliuk, N.M. Caffrey, T. Iakimov, V. Khranovskyy, I.A. Abrikosov, R. Yakimova. // Scientific Reports. 2017. V. 7, Article number: 3934.

5. D.A. Areshkin, D. Gunlycke, C.T. White. // Nano Lett. 2007, V. 7. P. 204-210.

6. C.T. White, J. Li, D. Gunlycke, J.W. Mintmire. // Nano Lett. 2007. V. 7. P. 825-830.

7. D. Gunlycke, C. White // Phys. Rev. B. 2008. V. 77. P. 115116.

8. Z.Yan et al. // J. Am. Chem. Soc. 2013. V. 135. P. 10755.

9. N.N. Konobeeva, M.B. Belonenko. // Physica B: Condensed.Matter, 2017. V. 514. P. 51-53.

10. N.N. Konobeeva, M.B. Belonenko. // Modern Physics Letters B. 2017. V. 31. P. 1750340.

11. N.N. Konobeeva // Journal of nano- and electronic physics. 2017. V. 9 (5). P. 05049.

12. M. A. Ruderman, C. Kittel // Phys. Rev. 1954. V. 96. P. 99.

13. T. Kasuya // Progr. Theor. Phys. 1956. V. 16. P. 45.

14. K. Yosida // Phys. Rev. 1957. V. 106. P. 893.

15. D.A. Abanin, A.V. Shytov, L.S. Levitov. arXiv: 1004.3678v2(2010).

16. V.V. Cheainov, O. Syljuasen, B.L. Altshuler, V.I. Fal'ko. arXiv:1002.2330v1 (2010).

17. B. Uchoa, T.G. Rappoport, A.H. Castro Neto. arXiv:1006.2512v1 (2010).

18. K. Szalowski // J. Phys.: Condens. Matter. 2013 V. 25. P. 166001
6